\documentclass[prd,twocolumn,showpacs,superscriptaddress,nofootinbib,floatfix,showkeys,10pt]{revtex4-2}
\usepackage{bm}
\usepackage{times}
\usepackage{braket}
\usepackage{amsfonts,amssymb,stmaryrd,latexsym,amsmath}
\usepackage[usenames,dvipsnames]{color}
\usepackage{slashed}
\usepackage{hyperref}
\usepackage{subfigure}
\usepackage{graphicx}
\usepackage{slashed}
\usepackage{orcidlink}
\usepackage{multirow}
\usepackage{appendix}
\usepackage{dashrule}
\usepackage{float}
\allowdisplaybreaks[1]

\definecolor{nicered}{rgb}{0.7,0.1,0.1}
\definecolor{nicegreen}{rgb}{0.1,0.5,0.1}
\definecolor{emph}{rgb}{1,0,0}
\definecolor{doub}{rgb}{0.7,0.2,1.0}
\definecolor{navyblue}{RGB}{0, 110, 184}

\hypersetup{colorlinks,citecolor=nicegreen,linkcolor=nicered,urlcolor=navyblue}

\begin{document}
	\title{Fully heavy tetraquark resonant states with different flavors} 
	\author{Wei-Lin Wu\,\orcidlink{0009-0009-3480-8810}}\email{wlwu@pku.edu.cn}
	\affiliation{School of Physics, Peking University, Beijing 100871, China}
	\author{Yao Ma\,\orcidlink{0000-0002-5868-1166}}\email{yaoma@pku.edu.cn}
	\affiliation{School of Physics and Center of High Energy Physics,
		Peking University, Beijing 100871, China}
	\author{Yan-Ke Chen\,\orcidlink{0000-0002-9984-163X}}\email{chenyanke@stu.pku.edu.cn}
	\affiliation{School of Physics, Peking University, Beijing 100871, China}
	\author{Lu Meng\,\orcidlink{0000-0001-9791-7138}}\email{lu.meng@rub.de}
	\affiliation{Institut f\"ur Theoretische Physik II, Ruhr-Universit\"at Bochum,  D-44780 Bochum, Germany }
	\author{Shi-Lin Zhu\,\orcidlink{0000-0002-4055-6906}}\email{zhusl@pku.edu.cn}
	\affiliation{School of Physics and Center of High Energy Physics,
		Peking University, Beijing 100871, China}
	
	\begin{abstract}
			We use the quark potential model to calculate the mass spectrum of the S-wave fully heavy tetraquark systems with different flavors, including the $ bc\bar b\bar c, bb\bar c\bar c, cc\bar c\bar b $ and $ bb\bar b\bar c $ systems. We employ the Gaussian expansion method to solve the four-body Schrödinger equation, and the complex scaling method to identify resonant states. The $ bc\bar b\bar c, bb\bar c\bar c, cc\bar c\bar b $ and $ bb\bar b\bar c $ resonant states are obtained in the mass regions of $ (13.2,13.5) $, $ (13.3,13.6) $, $ (10.0,10.3) $, $ (16.5,16.7) $ GeV, respectively. Among these states, the $ bc\bar b\bar c $ tetraquark states are the most promising ones to be discovered in the near future. We recommend the experimental exploration of the $ 1^{++} $ and $ 2^{++} $ $ bc\bar b\bar c $ states with masses near $ 13.3 $ GeV in the $ J/\psi\Upsilon $ channel. From the root-mean-square radii, we find that all the resonant states we have identified are compact tetraquark states.

	\end{abstract}
	
	\maketitle
	
	\section{Introduction}~\label{sec:intro}
	Hadron physics provides an excellent platform for studying the non-perturbative properties of quantum chromodynamics (QCD). In the past decades, tens of exotic hadrons beyond conventional mesons and baryons have been observed in experiments, which greatly advances the hadron spectroscopy. Many interpretations are proposed to understand these exotic states, including hadronic molecules, compact multiquark states, hybrid states, etc. More details can be found in recent reviews~\cite{Chen:2016qju,Hosaka:2016pey,Esposito:2016noz,Ali:2017jda,Lebed:2016hpi,Guo:2017jvc,Liu:2019zoy,Brambilla:2019esw,Meng:2022ozq,Chen:2022asf,Mai:2022eur}.
	
	Among various exotic hadrons, the fully heavy tetraquarks $ QQ\bar Q\bar Q\,(Q=b,c) $ have attracted great attention. Theoretically, they stand out as relatively clean systems, less affected by the creation and annihilation of the light quarks. In the absence of the long-range light meson exchange mechanism, the interactions between heavy quarks are dominated by the short-range gluon exchange and confinement. Therefore, the fully heavy tetraquark systems might have a tendency to form compact tetraquark states. Experimentally, great efforts and progress have been made in the search for the fully heavy tetraquark states. In the fully bottomed sector, $ bb\bar b\bar b $ candidates were searched for by the CMS~\cite{CMS:2016liw,CMS:2020qwa} and LHCb~\cite{LHCb:2018uwm} collaborations, but no significant signal was found. In the fully charmed sector, the LHCb discovered the first fully charmed tetraquark candidate
	$ X(6900) $~\cite{LHCb:2020bwg}, which was confirmed by the CMS~\cite{CMS:2023owd} and
	ATLAS~\cite{ATLAS:2023bft} collaborations. Meanwhile, more fully charmed tetraquark candidates were reported, including $ X(6600) $, $ X(7200) $ by the CMS~\cite{CMS:2023owd} and $ X(6400) $, $ X(6600) $, $ X(7200) $ by the ATLAS~\cite{ATLAS:2023bft}. Moreover, the CMS also observed triple $J/\psi$ production~\cite{CMS:2021qsn}, which may shed light on the future exploration of fully charmed hexaquarks.
	
	The existence of the fully charmed tetraquark candidates implies that similar tetraquark states may also exist in other fully heavy tetraquark systems, including the $ bc\bar b\bar c,bb\bar c\bar c, cc\bar c\bar b $ and $ bb\bar b\bar c $ systems. Among these systems, $ bc\bar b\bar c $ may be the most promising for experimental observation, since it only requires the production of two heavy quark-antiquark pairs. Some theoretical studies have been conducted on the fully heavy tetraquark systems with different flavors~\cite{Berezhnoy:2011xn,Wu:2016vtq,Chen:2016jxd,Richard:2017vry,Junnarkar:2018twb,Liu:2019zuc,Wang:2019rdo,Bedolla:2019zwg,Deng:2020iqw,Gordillo:2020sgc,Weng:2020jao,Yang:2021hrb,Hu:2022zdh,An:2022qpt,Mutuk:2022nkw,Zhang:2022qtp,Chen:2022mcr,Meng:2023jqk,Galkin:2023wox,Agaev:2024pej,Meng:2024yhu}. However, few works consider both compact diquark-antidiquark and molecular dimeson spatial configurations simultaneously and perform comprehensive dynamical calculations to identify genuine resonant states. In our previous work~\cite{Meng:2023jqk}, we incorporated both diquark-antidiquark and  dimeson spatial configurations, employing various quark models and few-body methods to conduct benchmark calculations for tetraquark bound states. Our results indicate that the Gaussian expansion method~\cite{HIYAMA2003223} is highly efficient in exploring tetraquark states, and that there are no bound states in the fully heavy tetraquark systems.
	
	In this study, we further investigate the S-wave fully heavy tetraquark resonant states with different flavors $ ( bc\bar b\bar c,bb\bar c\bar c, cc\bar c\bar b, bb\bar b\bar c ) $. We apply the complex scaling method~\cite{Aguilar:1971ve, Balslev:1971vb,Aoyama2006} to identify genuine
	resonant states from meson-meson scattering states. We employ the Gaussian expansion method~\cite{HIYAMA2003223} to solve the four-body Schrödinger equation, taking
	both diquark-antidiquark and dimeson spatial configurations into account. This framework has been successfully used to investigate the $ Qs\bar q\bar q $ and $ QQ\bar Q\bar Q $ $ (Q=b,c) $ systems~\cite{Chen:2023syh,Wu:2024euj}. For consistency, we adopt the AP1 quark potential model~\cite{Semay:1994ht, SilvestreBrac1996}, which was also used in our previous work on the fully heavy tetraquark $ QQ\bar Q\bar Q $ systems~\cite{Wu:2024euj}. Moreover, we analyze the spatial structures of the tetraquark states by calculating the root-mean-square radii, which allow us to distinguish between the compact and molecular tetraquark states. We also improve the numerical stability of the rms radii results from our previous work~\cite{Wu:2024euj}. This study may aid experimental exploration in the future. 
	
	This paper is organized as follows. In Sec.~\ref{sec:theo_framwork}, we introduce the theoretical framework, including the tetraquark Hamiltonian, the calculation methods, and the approach to analyzing the spatial structures. In Sec.~\ref{sec:result}, we present the numerical results and discuss properties of fully heavy tetraquark states with different flavors. In Sec.~\ref{sec:summary}, we give a summary of our findings.
 
	\section{Theoretical Framework}\label{sec:theo_framwork}
	
	\subsection{Hamiltonian}\label{subsec:hamiltonian}
	The nonrelativistic tetraquark Hamiltonian in the center-of-mass frame reads
	\begin{equation}
		H=\sum_{i=1}^4 (m_i+\frac{p_i^2}{2 m_i})+\sum_{i<j=1}^4 V_{ij
		},
	\end{equation}
	where the first two terms represent the mass and kinetic energy of the $i$-th (anti)quark and the last term represents the two-body interaction. In our previous study~\cite{Wu:2024euj}, we adopted three different quark potential models to study the fully charmed tetraquark systems and found that they give qualitatively consistent results. In this work, without prejudice to generality, we use the AP1 potential to study the fully heavy tetraquark systems with different flavors. The AP1 potential includes the one-gluon-exchange interaction and a $ 2/3 $ power quark confinement interaction, 
	\begin{equation}
		\label{eq:potential}
		\begin{aligned}
			V_{i j} =-\frac{3}{16} \boldsymbol\lambda_i \cdot \boldsymbol\lambda_j\left(-\frac{\kappa}{r_{i j}}+\lambda r_{i j}^{2/3}-\Lambda\right. \\
			\left.+\frac{8 \pi \kappa^{\prime}}{3 m_i m_j} \frac{\exp \left(-r_{i j}^2 / r_0^2\right)}{\pi^{3 / 2} r_0^3} \boldsymbol{S}_i \cdot \boldsymbol{S}_j\right),
		\end{aligned}
	\end{equation}
	where  $\boldsymbol\lambda_i$ is the $\mathrm{SU}(3)$ color Gell-Mann matrix, and $ \boldsymbol{S}_i $ is the spin operator.  The parameters of the AP1 model were determined by fitting the meson spectra, and we do not introduce any new free parameters. They are taken from Ref.~\cite{SilvestreBrac1996} and listed in Table~\ref{tab:para}. The theoretical masses of the heavy mesons as well as their root-mean-square (rms) radii are listed in Table~\ref{tab:meson}. It can be seen that the theoretical masses are in accordance with the experimental ones up to tens of MeV. We expect the errors for the tetraquark states to be of the same order.
	\begin{table*}[htbp]
		\centering
		\caption{The parameters in the AP1 quark potential model.}
		\label{tab:para}
		\begin{tabular*}{\hsize}{@{}@{\extracolsep{\fill}}cccccccc@{}}
			\hline\hline
			$ \kappa $ &$ \lambda { [\mathrm{GeV}^{5/3}]}$&$ \Lambda {\rm [GeV]} $&$ \kappa^\prime $&$ m_c {\rm [GeV]}$&$ m_b {\rm [GeV]}$&$ r_{0c} {\rm [GeV^{-1}]}$&$ r_{0b} {\rm [GeV^{-1}]}$\\
			\hline
			0.4242&0.3898&1.1313&1.8025&1.8190&5.206&1.2583&0.8928\\
			\hline\hline
		\end{tabular*}
	\end{table*}
	
	\begin{table}
		\centering
		\caption{The theoretical masses (in  $\mathrm{MeV}$) of heavy mesons, compared with the experimental results taken from Ref.~\cite{ParticleDataGroup:2022pth}. The theoretical rms radii (in fm) are listed in the last column. }
		\label{tab:meson}
		\begin{tabular*}{\hsize}{@{}@{\extracolsep{\fill}}cccc@{}}
			\hline\hline
			Mesons& $ m_{\rm Exp.} $&$ m_{\rm Theo.} $ &$ r^{\rm rms}_{\rm Theo.} $ \\
			\hline
			$\eta_c $ &2984&  2982&0.35\\
			$\eta_c(2S) $ &3638& 3605&0.78\\
			$\eta_c(3S) $ &-&  3986&1.15\\
			$J/\psi $ &3097&  3102&0.40\\
			$\psi(2S) $ &3686& 3645&0.81\\
			$\psi(3S) $&4039&  4011&1.17\\
			$\eta_b $ &9399& 9401 &0.20\\
			$\eta_b(2S) $&9999 & 10000&0.48\\
			$\eta_b(3S) $&-& 10326 &0.73\\
			$\Upsilon $ &9460&  9461 &0.21\\
			$\Upsilon(2S) $&10023 & 10014 &0.49\\
			$\Upsilon(3S) $&10355& 10335 &0.74\\
			$B_c $ &6274& 6269 &0.30\\
			$B_c(2S) $&6871 & 6854 &0.66\\
			$B_c^* $ &6329&  6338 &0.32\\
			$B_c^*(2S) $&- & 6875 &0.68\\
			
			\hline\hline
		\end{tabular*}
	\end{table}
	
	\subsection{Calculation methods}\label{subsec:the_calculation_method}
	To obtain possible bound and resonant states, we apply the complex scaling method (CSM). In the CSM~\cite{Aguilar:1971ve, Balslev:1971vb,Aoyama2006}, the coordinate $ \boldsymbol{r} $ and its conjugate momentum $ \boldsymbol{p} $ are transformed as
	\begin{equation}
		U(\theta) \boldsymbol{r}=\boldsymbol{r} e^{i \theta}, \quad U(\theta) \boldsymbol{p}=\boldsymbol{p} e^{-i \theta}.
	\end{equation}
	Under such a transformation, the complex-scaled Hamiltonian is no longer hermitian, which can be written as
	\begin{equation}
		H(\theta)=\sum_{i=1}^4 (m_i+\frac{p_i^2e^{-2i\theta}}{2 m_i})+\sum_{i<j=1}^4 V_{ij}(r_{ij}e^{i\theta}).
	\end{equation}
	According to the ABC theorem~\cite{Aguilar:1971ve,Balslev:1971vb}, the eigenenergies of the scattering states, bound states and resonant states can be obtained by solving the complex-scaled Schrödinger equations. The scattering states line up along rays starting from threshold energies with $ \operatorname{Arg}(E)=-2\theta $. The bound states are located on the negative real axis in the energy plane. The resonant states with mass $ M_R $ and width $ \Gamma_R $ can be detected at $ E_R=M_R-i\Gamma_R/2 $ when the complex scaling angle $ 2\theta>\left|\operatorname{Arg}(E_R)\right| $. Both the bound states and the resonant states remain stable as $ \theta $ changes.  
	
\begin{figure}[htbp]
		\centering
		\includegraphics[width=.9\linewidth]{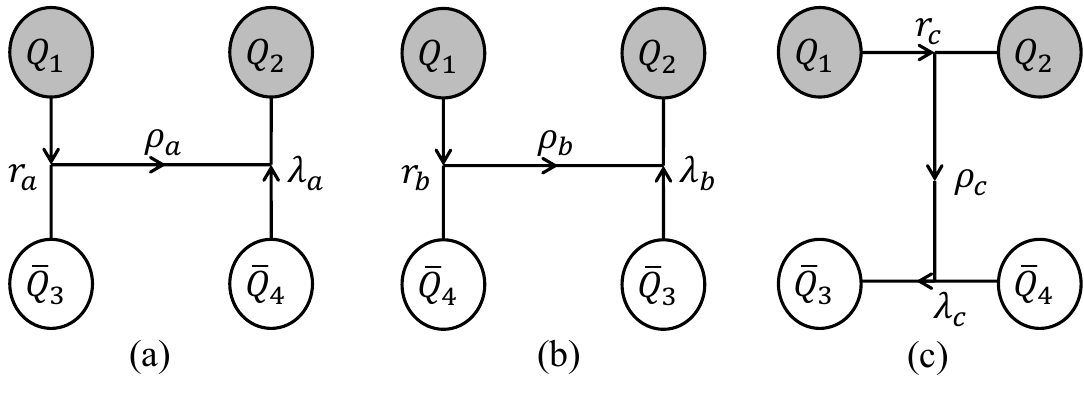}
		\caption{The Jacobian coordinates for two types of spatial configurations: (a), (b) for the dimeson configurations, and (c) for the diquark-antidiquark configuration.}
		\label{fig:jac}
	\end{figure}

	To solve the complex-scaled four-body Schrödinger equation, we apply the Gaussian expansion method (GEM)~\cite{HIYAMA2003223}. The wave functions of the S-wave tetraquark states with total angular momentum $ J $ are expanded as
	
	\begin{equation}\label{eq:wavefunction}
		\begin{aligned}
			\Psi^{J}(\theta)=\mathcal{A}\sum_{\rm{jac}}\sum_{\alpha,n_{i}}\,C^{(\rm{jac})}_{\alpha,n_{i}}(\theta)\chi_\alpha^J\phi_{n_{1}}(r_{jac})\phi_{n_{2}}(\lambda_{jac})\phi_{n_{3}}(\rho_{jac}),
		\end{aligned}
	\end{equation}
	where $ \mathcal{A} $ is the antisymmetric operator of identical particles. We consider three sets of spatial configurations (dimeson and diquark-antidiquark), which are denoted by $ (\rm{jac}) = (a),\,(b),\,(c) $. In each configuration, there are three independent Jacobian coordinates $r_{jac}, \lambda_{jac}$, $\rho_{jac}$, as shown in Fig.~\ref{fig:jac}. The spatial wave function $ \phi_{n_i}(r) $ takes the Gaussian form,
	\begin{equation}
		\begin{array}{c}
			\phi_{n_i}(r)=N_{n_{i}}e^{-\nu_{n_i}r^2},\\
			\nu_{n_i}=\nu_{1}\gamma^{n_i-1},
		\end{array}
	\end{equation}
	where $ N_{n_i} $ is the normalization factor. For the color-spin wave function $ \chi_\alpha^J $, we choose a complete set of basis given by
	\begin{equation}\label{eq:colorspin_wf1}
		\begin{aligned}
			\chi^{J}_{\bar 3_c\otimes 3_c,s_1,s_2}=\left[\left(Q_1Q_2\right)_{\bar 3_c}^{s_1}\left(\bar Q_3\bar Q_4\right)_{3_c}^{s_2}\right]_{1_c}^{J},\\
			\chi^{J}_{6_c\otimes \bar 6_c,s_1,s_2}=\left[\left(Q_1Q_2\right)_{6_c}^{s_1}\left(\bar Q_3\bar Q_4\right)_{\bar{6}_c}^{s_2}\right]_{1_c}^{J},\\
		\end{aligned}
	\end{equation}
	for all possible combinations of $ s_1, s_2, J $. 
	Finally, the expansion coefficients $ C^{(\rm{jac})}_{\alpha,n_i}(\theta) $ are determined by solving the energy eigenvalue equation,
	\begin{equation}
		H(\theta)\Psi^J(\theta)=E(\theta)\Psi^J(\theta).
	\end{equation}

	\subsection{Spatial structures}\label{subsec:spatial_distribution}

	The quark model does not make a priori assumptions about the structures of multiquark states, allowing both
	compact and molecular states. The root-mean-square (rms) radius is a commonly used criterion to distinguish between the compact and molecular tetraquark states. In our previous works~\cite{Chen:2023syh,Wu:2024euj}, we argued that the rms radii calculated using the complete wave functions could be misleading due to the antisymmetrization of identical particles. In order to eliminate the ambiguity arising from antisymmetrization, we proposed a new approach to calculate the rms radii. For systems with no identical particle ($ bc\bar b\bar c $), such ambiguity does not exist and we can calculate the rms radii using the complete wave function directly. For systems with one pair of identical particles ($cc\bar c\bar b, bb\bar b\bar c $), we decompose the complete antisymmetric wave function as 
	\begin{equation}\label{eq:wf_decompose1}
		\begin{aligned}
			\Psi^{J}(\theta)=&[(Q_1\bar Q_3)_{1_c}(Q_2\bar Q'_4)_{1_c}]_{1_c}\otimes|\psi_1(\theta)\rangle\\
			&+[(Q_2\bar Q_3)_{1_c}(Q_1\bar Q'_4)_{1_c}]_{1_c}\otimes|\psi_2(\theta)\rangle\\
			=&\mathcal{A}\,[(Q_1\bar Q_3)_{1_c}(Q_2\bar Q'_4)_{1_c}]_{1_c}\otimes|\psi_1(\theta)\rangle\\
			\equiv&\mathcal{A}\,\Psi^{J}_{nA}(\theta).
		\end{aligned}
	\end{equation}
	For systems with two pairs of identical particles $ (bb\bar c\bar c) $, we decompose the complete antisymmetric wave function as
	\begin{equation}\label{eq:wf_decompose2}
		\begin{aligned}
			\Psi^J(\theta)=&\sum_{s_1\geq s_2}\left([(Q_1\bar Q'_3)^{s_1}_{1_c}(Q_2\bar Q'_4)^{s_2}_{1_c}]^{J}_{1_c}\otimes|\psi_1^{s_1s_2}(\theta)\rangle\right.\\
			&+[(Q_1\bar Q'_3)^{s_2}_{1_c}(Q_2\bar Q'_4)^{s_1}_{1_c}]^{J}_{1_c}\otimes|\psi_2^{s_1s_2}(\theta)\rangle\\
			&+[(Q_1\bar Q'_4)^{s_1}_{1_c}(Q_2\bar Q'_3)^{s_2}_{1_c}]^{J}_{1_c}\otimes|\psi_3^{s_1s_2}(\theta)\rangle\\
			&+\left.[(Q_1\bar Q'_4)^{s_2}_{1_c}(Q_2\bar Q'_3)^{s_1}_{1_c}]^{J}_{1_c}\otimes|\psi_4^{s_1s_2}(\theta)\rangle\right)\\
			=&\mathcal{A}\sum_{s_1\geq s_2}[(Q_1\bar Q'_3)^{s_1}_{1_c}(Q_2\bar Q'_4)^{s_2}_{1_c}]^{J}_{1_c}\otimes|\psi_1^{s_1s_2}(\theta)\rangle\\
			\equiv&\mathcal{A}\,\Psi^{J}_{nA}(\theta),
		\end{aligned}
	\end{equation}
	where $ s_1, s_2 $ sum over spin configurations with total angular momentum $ J $. Instead of using the complete wave function $ \Psi^J(\theta) $, we use the non-antisymmetric component $ \Psi_{\mathrm{nA}}^J(\theta) $ to define the rms radius:
	\begin{equation}\label{eq:rmsr}
		r^{\mathrm{rms}}_{ij}\equiv \mathrm{Re}\left[\sqrt{\frac{\langle\Psi_{\mathrm{nA}}^J(\theta) | r_{ij}^2 e^{2i\theta}|\Psi_{\mathrm{nA}}^J(\theta)\rangle}{\langle\Psi_{\mathrm{nA}}^J(\theta) | \Psi_{\mathrm{nA}}^J(\theta)\rangle}}\right].
	\end{equation}
	
	The new definition of the rms radius is useful for investigating the spatial structures of the tetraquark states.  For a hadronic molecular state, $ r^{\mathrm{rms}}_{13} $ and $ r^{\mathrm{rms}}_{24} $ are expected to be the sizes of the constituent mesons, and much smaller than the other rms radii. For a compact tetraquark state, all rms radii in the four-body system should be of the same order. More discussions of the rms radii can be found in Refs.~\cite{Chen:2023syh,Wu:2024euj}.
	
	It should be emphasized that the inner products in the CSM are defined using the c-product~\cite{ROMO1968617}, 
	\begin{equation}
		\langle\phi_n \mid \phi_m\rangle\equiv\int \phi_n(r)\phi_{m}(r)d^3r,
	\end{equation}
	where the square of the wave function rather than the square of its magnitude is used. The rms radius calculated by the c-product is generally not real, but its real part can still reflect the internal quark clustering behavior if the width of the resonant state is not too large, as discussed in Ref.~\cite{homma1997matrix}.
	
	\section{Results and Discussions}\label{sec:result}
	We investigate the S-wave fully heavy tetraquark systems with different flavors, including the $ bc\bar b\bar c, bb\bar c\bar c, cc\bar c\bar b $ and $ bb\bar b\bar c $ systems. With the CSM, we calculate the complex energies of these systems. We choose varying complex scaling angles $ \theta $ to identify genuine resonant states. The meson-meson scattering states rotate along the continuum lines starting from the threshold energies, while bound states and resonant states do not shift with $ \theta $. We obtain a series of resonant states in all these systems, but no bound state exists below the lowest threshold. For convenience, we label the tetraquark resonant states obtained in our calculations as $ T_{Q_1Q_2\bar Q_3\bar Q_4,J^{P(C)}}(M) $, where $ Q_1Q_2\bar Q_3\bar Q_4 $ is the quark content and $ M $ is the mass of the state.
	
	\subsection{$ bc\bar b\bar c $}
	
	\begin{figure*}[tbp]
		\centering
		\includegraphics[width=1\linewidth]{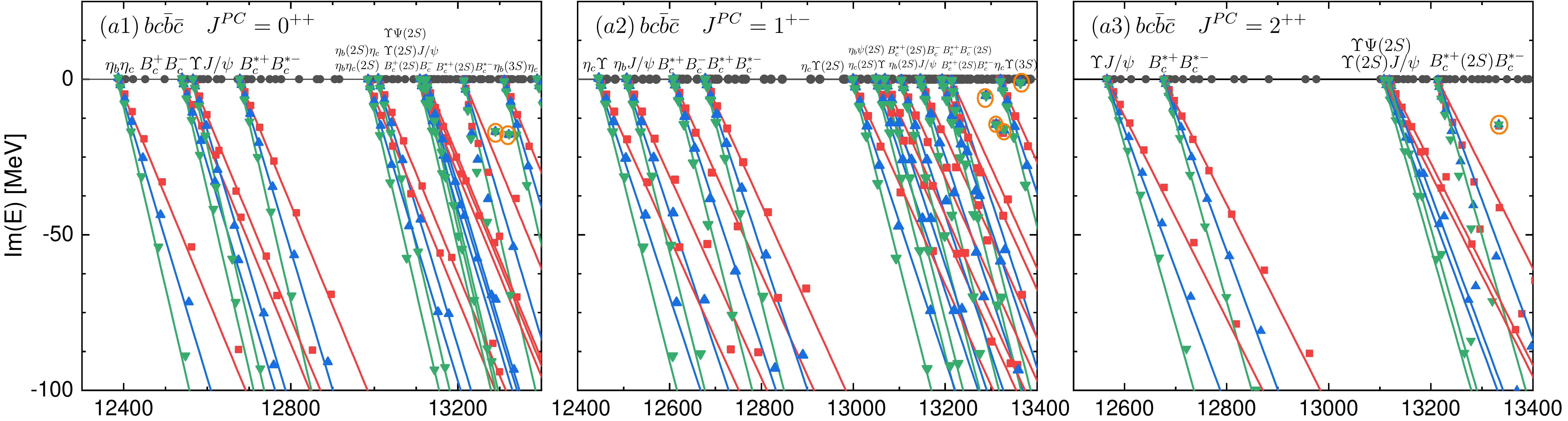}
		\includegraphics[width=1\linewidth]{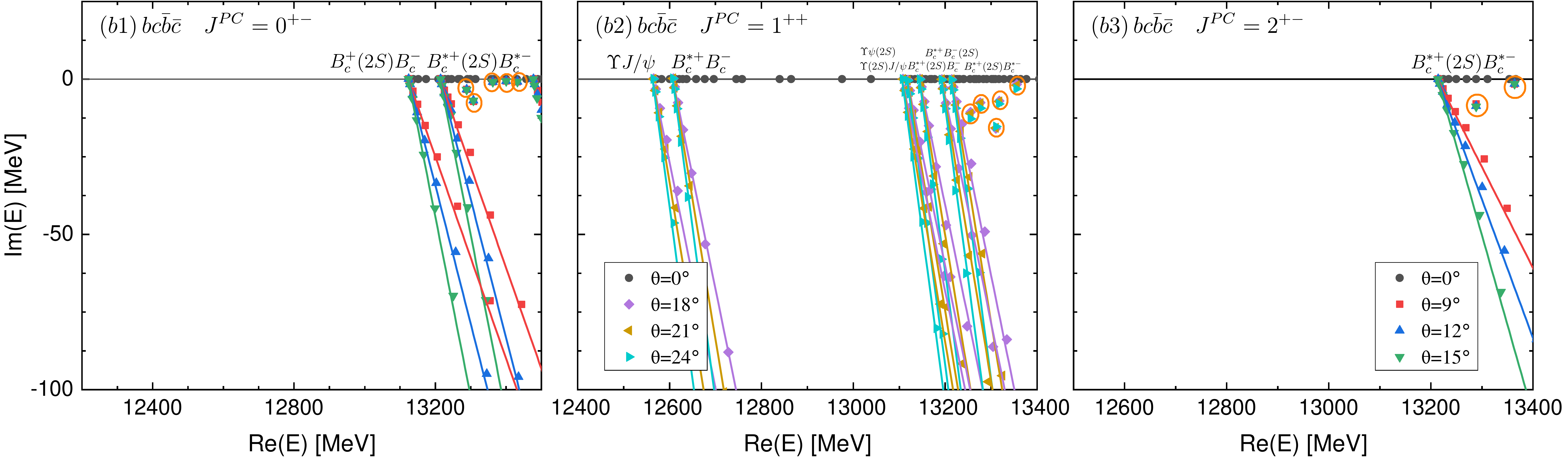}
		\caption{The complex energy eigenvalues of the  $bc\bar b\bar c$ states with varying $\theta$ in the CSM. The solid lines represent the continuum lines rotating along $\operatorname{Arg}(E)=-2 \theta$. The resonant states do not shift as $\theta$ changes and are highlighted by the orange circles.}
		\label{fig:bcbc}
	\end{figure*}
	\begin{figure*}[htbp]
		\centering
		\includegraphics[width=1\linewidth]{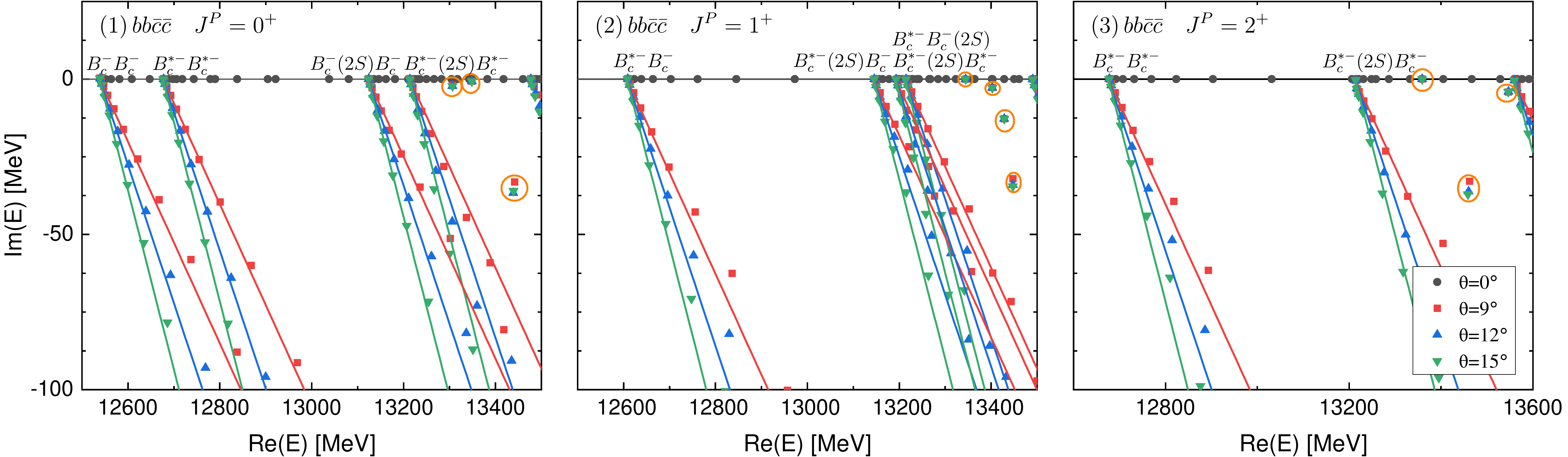}
		\caption{The complex energy eigenvalues of the  $bb\bar c\bar c$ states with varying $\theta$ in the CSM. The solid lines represent the continuum lines rotating along $\operatorname{Arg}(E)=-2 \theta$. The resonant states do not shift as $\theta$ changes and are highlighted by the orange circles.}
		\label{fig:bbcc}
	\end{figure*}
	\begin{table*}[htbp]
		\centering
		\caption{The complex energies (in MeV), the proportions of different color configurations and the rms radii (in fm) of the $ bc\bar b\bar c $ resonant states. }
		\label{tab:bcbc_structure}
		\begin{tabular*}{\hsize}{@{}@{\extracolsep{\fill}}cccccccc@{}}
			\hline\hline
			$ J^{PC}$& $ M-i\Gamma/2 $ & $ \chi_{\bar{3}_c\otimes3_c} $ &$ \chi_{6_c\otimes \bar6_c} $& $ r_{b\bar{b}}^{\mathrm{rms}} $&$ r_{c\bar{c}}^{\mathrm{rms}} $&$ r_{b\bar{c}}^{\mathrm{rms}} = r_{c\bar{b}}^{\mathrm{rms}} $&$ r_{bc}^{\mathrm{rms}} = r_{\bar{b}\bar{c}}^{\mathrm{rms}} $\\
			\hline
			$ 0^{++} $&$ 13290-17i $&$ 56\% $&$ 44\% $&$ 0.58 $&$ 0.73 $&$ 0.46 $&$ 0.57 $\\
			&$ 13322-18i $&$ 56\% $&$ 44\% $&$ 0.38 $&$ 0.63 $&$ 0.65 $&$ 0.48$\\	
			$ 1^{+-} $&$ 13289-5i $&$ 48\% $&$ 52\% $&$ 0.32 $&$ 0.71 $&$ 0.60 $&$ 0.61 $\\
			&$ 13311-15i $&$ 53\% $&$ 47\% $&$ 0.50 $&$ 0.70 $&$ 0.53 $&$ 0.58 $\\
			&$ 13328-16i $&$ 54\% $&$ 46\% $&$ 0.30 $&$ 0.59 $&$ 0.60 $&$ 0.50 $\\
			&$ 13364-1i $&$ 49\% $&$ 51\% $&$ 0.43 $&$ 0.58 $&$ 0.56 $&$ 0.56 $\\
			$ 2^{++} $&$ 13333-14i $&$ 53\% $&$ 47\% $&$ 0.44 $&$ 0.68 $&$ 0.53 $&$ 0.53 $\\
			$ 0^{+-} $&$ 13289-3i $&$ 47\% $&$ 53\% $&$ 0.32 $&$ 0.70 $&$ 0.60 $&$ 0.61 $\\
			&$ 13308-7i $&$ 46\% $&$ 54\% $&$ 0.36 $&$ 0.52 $&$ 0.54 $&$ 0.49 $\\
			&$ 13362-1i $&$ 50\% $&$ 50\% $&$ 0.42 $&$ 0.58 $&$ 0.56 $&$ 0.55$\\
			&$ 13400-1i $&$ 67\% $&$ 33\% $&$ 0.41 $&$ 0.59 $&$ 0.53 $&$ 0.56$\\
			&$ 13432-1i $&$ 64\% $&$ 36\% $&$ 0.43 $&$ 0.61 $&$ 0.54 $&$ 0.58$\\
			
			$ 1^{++} $&$ 13255-11i $&$ 35\% $&$ 65\% $&$ 0.32 $&$ 0.70 $&$ 0.60 $&$ 0.60 $\\
			&$ 13276-8i $&$ 45\% $&$ 55\% $&$ 0.31 $&$ 0.70 $&$ 0.59 $&$ 0.60 $\\
			&$ 13310-16i $&$ 56\% $&$ 44\% $&$ 0.50 $&$ 0.71 $&$ 0.52 $&$ 0.57 $\\
			&$ 13318-7i $&$ 48\% $&$ 52\% $&$ 0.41 $&$ 0.55 $&$ 0.55 $&$ 0.53 $\\
			&$ 13355-3i $&$ 45\% $&$ 55\% $&$ 0.41 $&$ 0.56 $&$ 0.54 $&$ 0.54 $\\
			$ 2^{+-} $&$ 13289-9i $&$ 41\% $&$ 59\% $&$ 0.57 $&$ 0.85 $&$ 0.61 $&$ 0.78 $\\
			&$ 13364-2i $&$ 45\% $&$ 55\% $&$ 0.42 $&$ 0.58 $&$ 0.56 $&$ 0.56 $\\
			\hline\hline
		\end{tabular*}
	\end{table*}
	The $ bc\bar b\bar c $ tetraquark is a neutral system with definite charge parity. For the S-wave neutral tetraquark systems, possible quantum numbers include $ J^{PC}=0^{++},1^{+-},2^{++},0^{+-},1^{++},2^{+-} $. In Ref.~\cite{Wu:2024euj}, we introduced a method to determine the C-parity of the neutral tetraquark
	states by decomposing the Hilbert space. 
	
	The complex energies of the $ bc\bar b\bar c $ systems are shown in Fig~\ref{fig:bcbc}. We obtain a series of resonant states, whose complex energies, proportions of different color configurations and rms radii are summarized in Table~\ref{tab:bcbc_structure}. The $ bc\bar b\bar c $ resonant states are located in the mass region $ (13.2,13.5) $ GeV. The different rms radii of these states are of the same order, approximately matching the sizes of the corresponding 2S mesons. This indicates that they are compact tetraquark states. 
	
\begin{figure}[H]
		\centering
		\includegraphics[width=.85\linewidth]{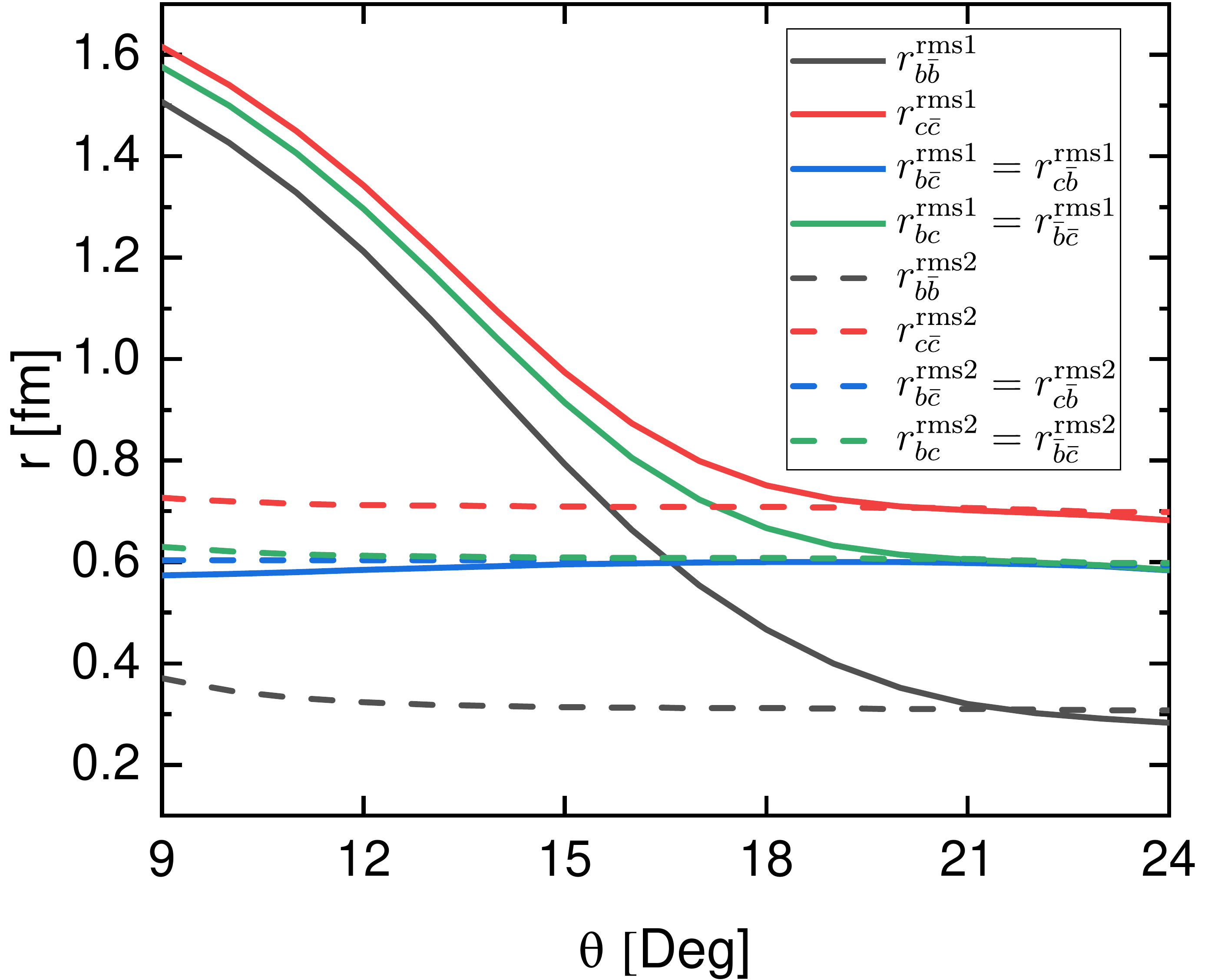}
		\caption{The rms radii (in fm) of $ T_{bc\bar b\bar c,1^{++}}(13255) $ (solid lines) and $ T_{bc\bar b\bar c,1^{+-}}(13289) $ (dashed lines) calculated using different complex scaling angles.}
		\label{fig:rtheta}
	\end{figure}

	In order to obtain good numerical convergence of the rms radii of the resonant states, it is important to choose appropriate complex scaling angles $ \theta $ so that the continuum lines are not too close to the resonant states, as discussed in Ref.~\cite{homma1997matrix}. Otherwise, the rms radii results could be affected by the scattering states, becoming numerically unstable and potentially leading to false conclusions. To illustrate this, we use different angles to calculate the rms radii of $ T_{bc\bar b\bar c,1^{++}}(13255) $ and $ T_{bc\bar b\bar c,1^{+-}}(13289) $, which are respectively denoted by $ r^{\rm rms1} $ and $ r^{\rm rms2} $ and shown in Fig.~\ref{fig:rtheta}. The values of $ r^{\rm rms1} $ are numerically unstable for $ \theta=9^\circ\sim 15^\circ $, since the state $ T_{bc\bar b\bar c,1^{++}}(13255) $ is located too close to the $ B_c^{*+}(2S)B_c^{*-} $ continuum line for these angles. Therefore, we need to choose larger angles $ (\theta=21^\circ,24^\circ) $ for the $ 1^{++} $ system to obtain convergent rms radii results. But for other resonant states such as $ T_{bc\bar b\bar c,1^{+-}}(13289) $, we can see from Fig.~\ref{fig:bcbc} that they are not located close to the continuum lines. As a result, we can obtain good convergent values of $ r^{\rm rms2} $ using $ \theta=9^\circ\sim15^\circ $.
	
	We also use the same method to improve our previous rms radii results of the fully charmed tetraquark states~\cite{Wu:2024euj}. In our previous work, we suggested that three $ cc\bar c\bar c $ resonant states, including two $ X(7200) $ candidates, might have a molecular configuration, but their rms radii were less numerically accurate. By choosing larger angles, we obtain more numerically stable results, which are compared with the previous ones in Table~\ref{tab:cccc}. With the improved results, we find that these three states actually have compact tetraquark configuration.
	
	\begin{table*}[htbp]
		\centering 
		\caption{The previous (P.) and improved (I.) results for the complex energies (in MeV) and rms radii (in fm) of the $ cc\bar c\bar c $ resonant states in Ref.~\cite{Wu:2024euj}. The previous results are taken from Ref.~\cite{Wu:2024euj}. The last column shows the spatial configurations of the states, where C. and M. represent the compact tetraquark and molecular configurations, respectively.}
		\label{tab:cccc}
		\begin{tabular*}{\hsize}{@{}@{\extracolsep{\fill}}cccccccc@{}}
			\hline\hline
			&$ J^{PC} $& $ M-i\Gamma/2 $ & $ r_{c_1\bar{c}_3}^{\mathrm{rms}} $&$ r_{c_2\bar{c}_4}^{\mathrm{rms}} $&$ r_{c_1\bar{c}_4}^{\mathrm{rms}} $=$ r_{c_2\bar{c}_3}^{\mathrm{rms}} $&$ r_{c_1c_2}^{\mathrm{rms}} $=$ r_{\bar{c}_3\bar{c}_4}^{\mathrm{rms}} $&Configurations\\
			\hline
			P.&$ 0^{++} $&$ 7173-20i $&$0.89$&$0.89$&$2.31$&$2.28$&M.\\
			I.&&$ 7167-19i $&$ 0.91 $&$ 0.91 $&$ 0.90$&$ 0.67 $&C.\\
			P.&$ 1^{+-} $&$ 7191-32i $&$0.71$&$1.08$&$2.09$&$2.08$&M.\\
			I.&&$ 7181-27i $&$0.91$&$0.93$&$0.87$&$0.61$&C.\\
			P.&$ 2^{++} $&$ 7214-30i $&$0.92$&$0.92$&$1.93$&$1.88$&M.\\
			I.&&$ 7204-29i $& $ 0.94 $&$ 0.94 $&$ 0.85 $&$ 0.62 $&C.\\
			\hline\hline
		\end{tabular*}
	\end{table*}
	
	Among all the fully heavy tetraquark systems with different flavors, the $ bc\bar b\bar c $ tetraquark is the most promising one to be discovered in experiments, since it only requires the production of two heavy quark-antiquark pairs.  Moreover, the $ 1^{++} $ and $ 2^{++} $ $ bc\bar b\bar c $ tetraquark resonant states, including the $ T_{bc\bar b\bar c,1^{++}}(13255)$, $T_{bc\bar b\bar c,1^{++}}(13276)$, $T_{bc\bar b\bar c,1^{++}}(13310)$, $T_{bc\bar b\bar c,1^{++}}(13318)$, $T_{bc\bar b\bar c,1^{++}}(13355)$ and $T_{bc\bar b\bar c,2^{++}}(13333) $, can decay into the $ J/\psi\Upsilon $ channel, which can be efficiently reconstructed in experiments. Therefore, we recommend experimental exploration of the $ 1^{++} $ and $ 2^{++} $ $ bc\bar b\bar c $ states with masses near $ 13.3 $ GeV in the $ J/\psi\Upsilon $ channel.

	\subsection{$ bb\bar c\bar c $}	
	The complex eigenenergies of the $ 0^+,1^+$ and $ 2^+ $ $ bb\bar c\bar c $ systems are shown in Fig.~\ref{fig:bbcc}. We obtain some resonant states, whose complex energies, proportions of different color configurations and rms radii are summarized in Table~\ref{tab:bbcc_structure}. The $ bb\bar c\bar c $ resonant states are located in the mass region $ (13.3,13.6) $ GeV.  The different rms radii of these states are of the same order, indicating that they are compact tetraquark states.
	\begin{figure}[htbp]
		\centering
		\includegraphics[width=.85\linewidth]{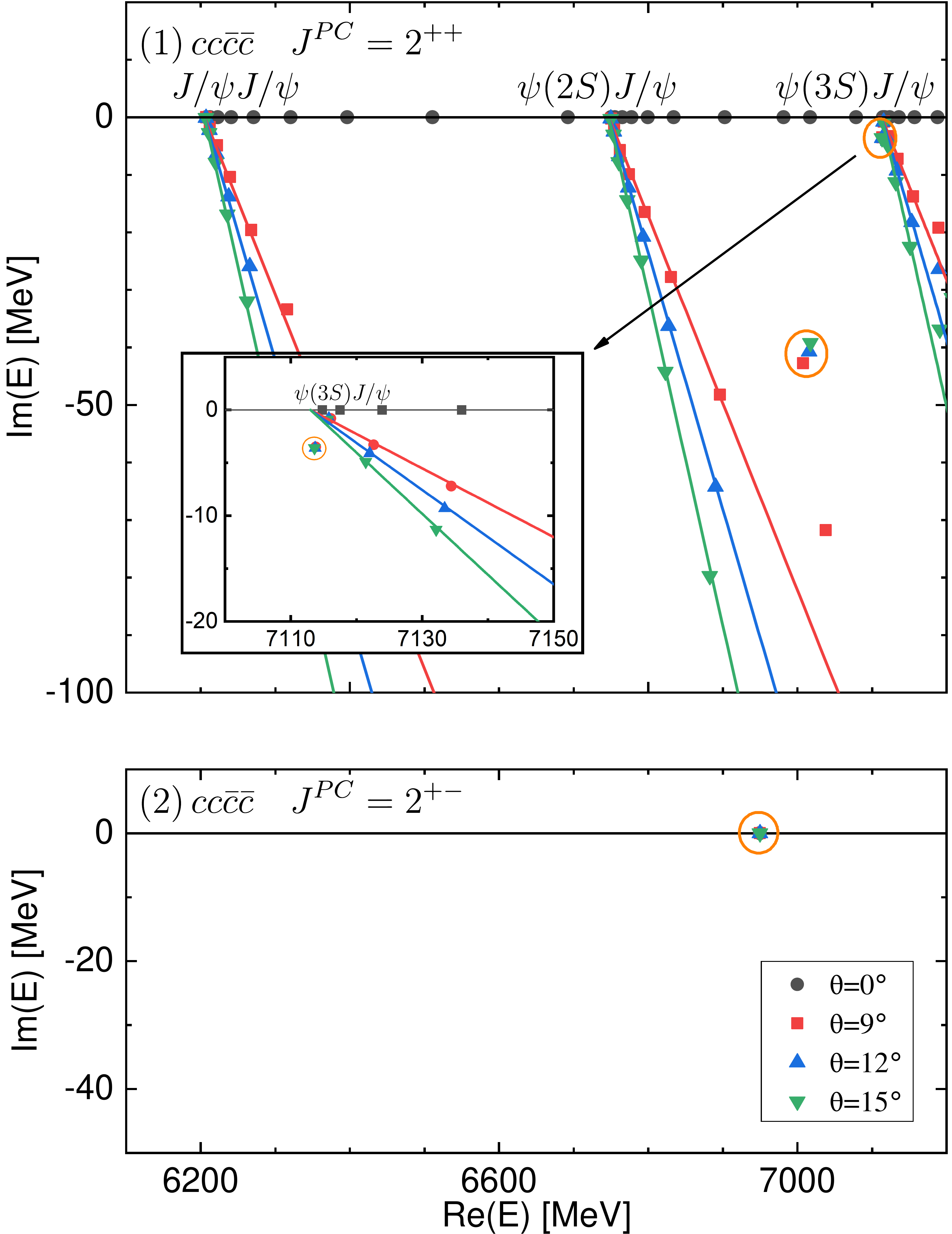}
		\caption{The complex energy eigenvalues of the $ 2^+ $ $cc\bar c\bar c$ states with varying $\theta$ in the CSM. The solid lines represent the continuum lines rotating along $\operatorname{Arg}(E)=-2 \theta$. The resonant states do not shift as $\theta$ changes and are highlighted by the orange circles.}
		\label{fig:4cJ2}
	\end{figure}
	
	Unlike the $ cc\bar c\bar c $ and $ bb\bar b\bar b $ systems, the $ bb\bar c\bar c $ system does not have definite C-parity. However, when compared with our previous work on the fully heavy tetraquark $ QQ\bar Q\bar Q\,(Q=b,c)$ systems~\cite{Wu:2024euj}, we still find some similarities in the $ bb\bar c\bar c $ system. For example, in the $ cc\bar c\bar c $ system with $ J^P=2^+ $, there exist two resonant states with positive C-parity and one zero-width state with negative C-parity below the $ \psi(3S)J/\psi $ threshold, as shown in Fig.~\ref{fig:4cJ2}. Similarly, in the $ 2^+ $ $ bb\bar c\bar c $ system, we obtain two resonant states with nonzero widths and one zero-width state below the  $ B_c^*(3S)B_c^* $ threshold. Comparing Fig.~\ref{fig:bbcc}.(3) and Fig.~\ref{fig:4cJ2}, it is evident that the $2^+$ $bb\bar c\bar c$ energy plot shares a similar pattern with the superposition of the $2^{++}$ and $2^{+-}$ $cc\bar c\bar c$ energy plots. Such similarities also exist in $ J^P=0^+ $ and $ 1^+ $ systems. In our previous work~\cite{Wu:2024euj}, we identified two resonant states with $ J^{PC}=0^{++} $ and $ 2^{++} $ as candidates of $ X(6900) $. Their analogs in $ bb\bar c\bar c $ systems are $ T_{bb\bar c\bar c,0^+}(13439) $ and $ T_{bb\bar c\bar c,2^+}(13460) $. 
	
	\begin{table*}[htbp]
		\centering
		\caption{The complex energies (in MeV), the proportions of different color configurations and the rms radii (in fm) of the $ bb\bar c\bar c $ resonant states. }
		\label{tab:bbcc_structure}
		\begin{tabular*}{\hsize}{@{}@{\extracolsep{\fill}}cccccccccc@{}}
			\hline\hline
			$ J^{P}$& $ M-i\Gamma/2 $ & $ \chi_{\bar{3}_c\otimes3_c} $ &$ \chi_{6_c\otimes \bar6_c} $& $ r_{b_1\bar{c}_1}^{\mathrm{rms}} $&$ r_{b_2\bar{c}_2}^{\mathrm{rms}} $&$ r_{b_1b_2}^{\mathrm{rms}} $&$ r_{\bar{c}_1\bar{c}_2}^{\mathrm{rms}} $&$ r_{b_1\bar{c}_2}^{\mathrm{rms}} $&$ r_{b_2\bar{c}_1}^{\mathrm{rms}} $\\
			\hline
			$ 0^{+} $&$ 13306-2i $&$ 35\% $&$ 65\% $&$ 0.50 $&$ 0.50 $&$ 0.53 $&$ 0.65 $&$ 0.56 $&$ 0.56 $\\
			&$ 13349-1i $&$ 70\% $&$ 30\% $&$ 0.51 $&$ 0.51 $&$ 0.50 $&$ 0.64 $&$ 0.55 $&$ 0.55 $\\
			&$ 13439-37i $& $ 89 \%$ &$11\% $&$ 0.68 $&$ 0.68 $&$ 0.28 $&$ 0.61 $&$ 0.65 $&$ 0.65 $\\
			$ 1^{+} $&$ 13344 $&$ 85\% $&$ 15\% $&$ 0.53 $&$ 0.49 $&$ 0.49 $&$ 0.62 $&$ 0.52 $&$ 0.54 $\\
			&$ 13402-3i $&$ 77\% $&$ 23\% $&$ 0.52 $&$ 0.55 $&$ 0.41 $&$ 0.59 $&$ 0.57 $&$ 0.57 $\\
			&$ 13429-13i $&$ 32\% $&$ 68\% $&$ 0.55 $&$ 0.57 $&$ 0.38 $&$ 0.62 $&$ 0.62 $&$ 0.61 $\\
			&$ 13448-34i $&$ 83\% $&$ 17\% $&$ 0.67 $&$ 0.68 $&$ 0.39 $&$ 0.68 $&$ 0.70 $&$ 0.69 $\\
			$ 2^{+} $&$ 13359 $&$ 86\% $&$ 14\% $&$ 0.52 $&$ 0.52 $&$ 0.49 $&$ 0.63 $&$ 0.54 $&$ 0.54 $\\
			&$13460-36i $& $ 82\% $&$ 18\% $&$ 0.67 $&$ 0.67 $&$ 0.47 $&$ 0.75 $&$ 0.73 $&$ 0.73 $\\
			&$ 13547-4i $&$ 80\% $&$ 20\% $&$ 0.76 $&$ 0.76 $&$ 0.50 $&$ 0.76 $&$ 0.80 $&$ 0.80 $\\
			
			\hline\hline
		\end{tabular*}
	\end{table*}

	\subsection{$ cc\bar c\bar b $ and $ bb\bar b\bar c $}\label{subsec:QQQQ'}

	\begin{figure*}[htbp]
		\centering
		\includegraphics[width=1\linewidth]{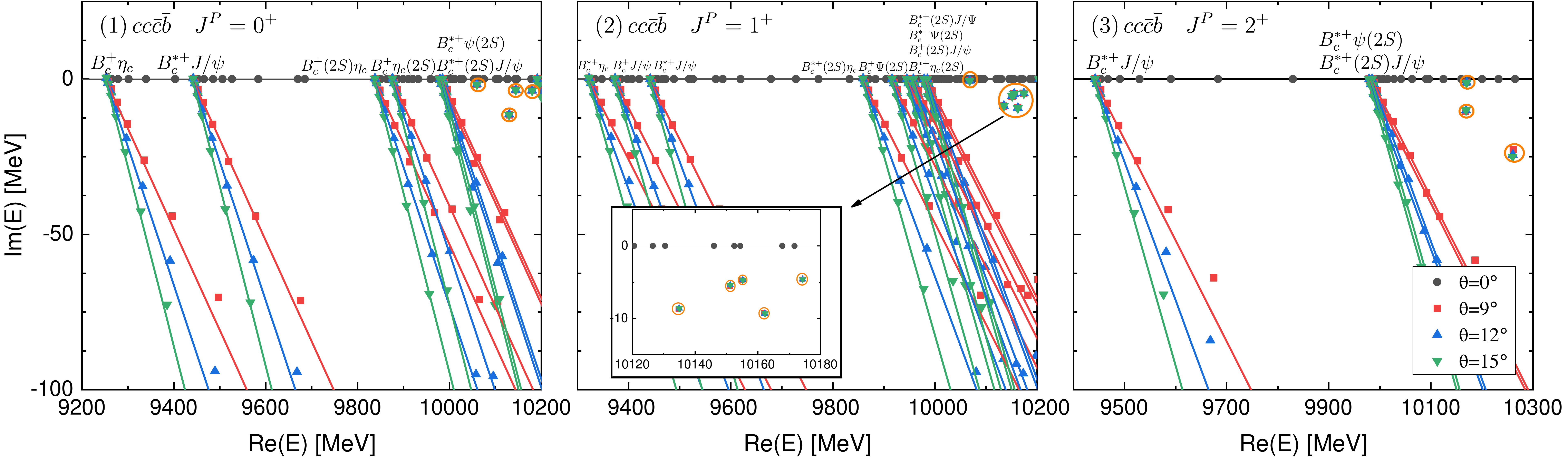}
		\caption{The complex energy eigenvalues of the  $cc\bar c\bar b$ states with varying $\theta$ in the CSM. The solid lines represent the continuum lines rotating along $\operatorname{Arg}(E)=-2 \theta$. The resonant states do not shift as $\theta$ changes and are highlighted by the orange circles.}
		\label{fig:cccb}
	\end{figure*}
	
	\begin{figure*}[htbp]
		\centering
		\includegraphics[width=1\linewidth]{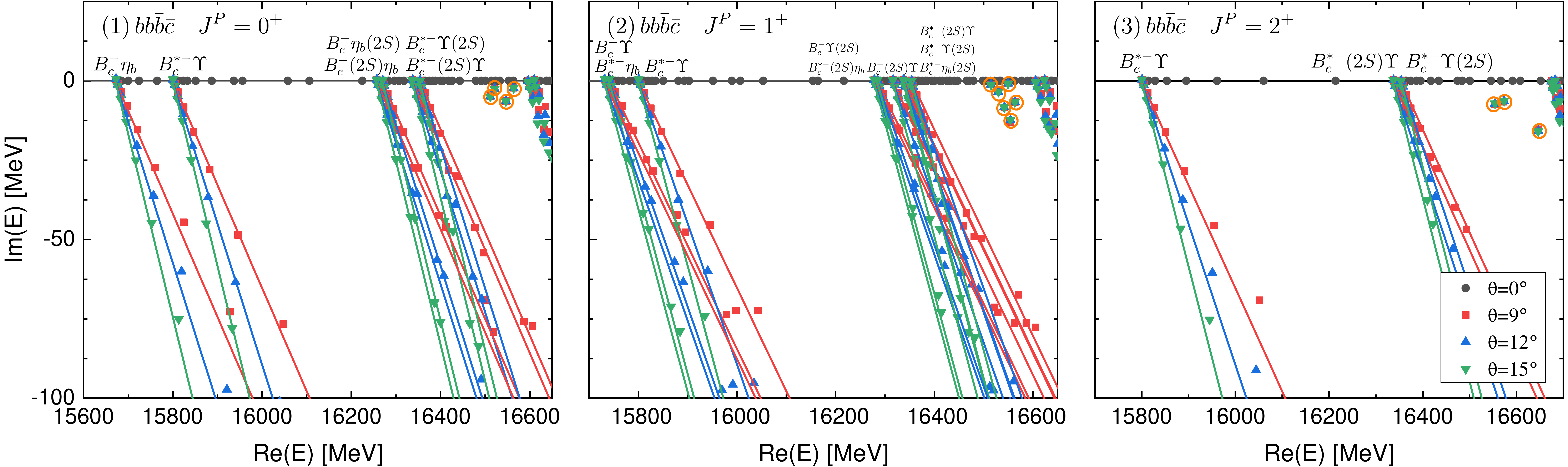}
		\caption{The complex energy eigenvalues of the  $bb\bar b\bar c$ states with varying $\theta$ in the CSM. The solid lines represent the continuum lines rotating along $\operatorname{Arg}(E)=-2 \theta$. The resonant states do not shift as $\theta$ changes and are highlighted by the orange circles.}
		\label{fig:bbbc}
	\end{figure*}

	\begin{table*}[htbp]
		\centering
		\caption{The complex energies (in MeV), the proportions of different color configurations and the rms radii (in fm) of the $ cc\bar c\bar b $ resonant states. }
		\label{tab:cccb_structure}
		\begin{tabular*}{\hsize}{@{}@{\extracolsep{\fill}}cccccccccc@{}}
			\hline\hline
			$ J^{P}$& $ M-i\Gamma/2 $ & $ \chi_{\bar{3}_c\otimes3_c} $ &$ \chi_{6_c\otimes \bar6_c} $&$ r_{c_1\bar{c}}^{\mathrm{rms}} $& $ r_{c_2\bar b}^{\mathrm{rms}} $&$ r_{c_1c_2}^{\mathrm{rms}} $&$ r_{\bar c\bar b}^{\mathrm{rms}} $&$ r_{{c}_1\bar b}^{\mathrm{rms}} $&$ r_{{c}_2\bar c}^{\mathrm{rms}} $\\
			\hline
			$ 0^{+} $&$ 10059-2i $&$ 44\% $&$ 56\% $&$ 0.70 $&$ 0.45 $&$ 0.67 $&$ 0.56 $&$ 0.70 $&$ 0.63 $\\
			&$ 10130-11i $&$ 65\% $&$ 35\% $&$ 0.59 $&$ 0.55 $&$ 0.66 $&$ 0.54 $&$ 0.55 $&$ 0.69 $\\
			&$ 10144-4i $& $ 81 \%$ &$19\% $&$ 0.72 $&$ 0.45 $&$ 0.57 $&$ 0.66 $&$ 0.58 $&$ 0.68 $\\
			&$ 10180-4i $& $ 43 \%$ &$57\% $&$ 0.61 $&$ 0.59 $&$ 0.71 $&$ 0.57 $&$ 0.59 $&$ 0.69 $\\
			$ 1^{+} $&$ 10068-0.5i $&$ 38\% $&$ 62\% $&$ 0.71 $&$ 0.45 $&$ 0.69 $&$ 0.57 $&$ 0.71 $&$ 0.64 $\\
			&$ 10135-8i $&$ 73\% $&$ 27\% $&$ 0.58 $&$ 0.56 $&$ 0.68 $&$ 0.51 $&$ 0.56 $&$ 0.66 $\\
			&$ 10151-5i $&$ 76\% $&$ 24\% $&$ 0.78 $&$ 0.31 $&$ 0.46 $&$ 0.72 $&$ 0.53 $&$ 0.69 $\\
			&$ 10155-5i $&$ 92\% $&$ 8\% $&$ 0.65 $&$ 0.54 $&$ 0.67 $&$ 0.58 $&$ 0.57 $&$ 0.66 $\\
			&$ 10162-9i $&$ 84\% $&$ 16\% $&$ 0.64 $&$ 0.56 $&$ 0.67 $&$ 0.56 $&$ 0.56 $&$ 0.73 $\\
			&$ 10174-4i $&$ 37\% $&$ 63\% $&$ 0.60 $&$ 0.59 $&$ 0.72 $&$ 0.55 $&$ 0.60 $&$ 0.69 $\\
			$ 2^{+} $&$ 10169-10i $&$ 75\% $&$ 25\% $&$ 0.70 $&$ 0.50 $&$ 0.58 $&$ 0.59 $&$ 0.53 $&$ 0.76 $\\
			&$ 10170-1i $&$ 95\% $&$ 5\% $&$ 0.65 $&$ 0.53 $&$ 0.68 $&$ 0.60 $&$ 0.57 $&$ 0.63 $\\
			&$10260-24i $& $ 84\% $&$ 16\% $&$ 0.77 $&$ 0.70 $&$ 0.67 $&$ 0.53 $&$ 0.72 $&$ 0.81 $\\
			\hline\hline
		\end{tabular*}
	\end{table*}
	\begin{table*}[htbp]
		\centering
		\caption{The complex energies (in MeV), the proportions of different color configurations and the rms radii (in fm) of the $ bb\bar b\bar c $ resonant states. }
		\label{tab:bbbc_structure}
		\begin{tabular*}{\hsize}{@{}@{\extracolsep{\fill}}cccccccccc@{}}
			\hline\hline
			$ J^{P}$& $ M-i\Gamma/2 $ & $ \chi_{\bar{3}_c\otimes3_c} $ &$ \chi_{6_c\otimes \bar6_c} $&$ r_{{b}_1\bar b}^{\mathrm{rms}} $& $ r_{{b}_2\bar c}^{\mathrm{rms}} $&$ r_{{b}_1{b}_2}^{\mathrm{rms}} $&$ r_{\bar b\bar c}^{\mathrm{rms}} $&$ r_{{b}_1\bar c}^{\mathrm{rms}} $&$ r_{{b}_2\bar b}^{\mathrm{rms}} $\\
			\hline
			$ 0^{+} $&$ 16511-5i $&$ 61\% $&$ 39\% $&$ 0.38 $&$ 0.51 $&$ 0.41 $&$ 0.52 $&$ 0.53 $&$ 0.51 $\\
			&$ 16521-2i $&$ 30\% $&$ 70\% $&$ 0.35 $&$ 0.55 $&$ 0.46 $&$ 0.58 $&$ 0.57 $&$ 0.42 $\\
			&$ 16546-6i $& $ 55 \%$ &$45\% $&$ 0.39 $&$ 0.54 $&$ 0.41 $&$ 0.53 $&$ 0.56 $&$ 0.49 $\\
			&$ 16563-2i $& $ 77 \%$ &$23\% $&$ 0.40 $&$ 0.51 $&$ 0.49 $&$ 0.48 $&$ 0.50 $&$ 0.42 $\\
			$ 1^{+} $&$ 16515-1i $&$ 28\% $&$ 72\% $&$ 0.40 $&$ 0.48 $&$ 0.48 $&$ 0.51 $&$ 0.51 $&$ 0.48 $\\
			&$ 16530-3i $&$ 64\% $&$ 36\% $&$ 0.34 $&$ 0.56 $&$ 0.40 $&$ 0.56 $&$ 0.57 $&$ 0.46 $\\
			&$ 16542-8i $&$ 46\% $&$ 54\% $&$ 0.34 $&$ 0.58 $&$ 0.35 $&$ 0.61 $&$ 0.61 $&$ 0.45 $\\
			&$ 16550-1i $&$ 67\% $&$33\% $&$ 0.36 $&$ 0.55 $&$ 0.46 $&$ 0.53 $&$ 0.54 $&$ 0.43 $\\
			&$ 16554-12i $&$ 59\% $&$ 41\% $&$ 0.42 $&$ 0.51 $&$ 0.41 $&$ 0.48 $&$ 0.54 $&$ 0.45 $\\
			&$ 16564-7i $&$ 100\% $&$ 0\% $&$ 0.44 $&$ 0.47 $&$ 0.50 $&$ 0.42 $&$ 0.48 $&$ 0.42 $\\
			$ 2^{+} $&$ 16554-7i $&$ 49\% $&$ 51\% $&$ 0.31 $&$ 0.61 $&$ 0.31 $&$ 0.64 $&$ 0.63 $&$ 0.42 $\\
			&$ 16574-6i $&$ 95\% $&$ 5\% $&$ 0.45 $&$ 0.47 $&$ 0.50 $&$ 0.41 $&$ 0.48 $&$ 0.42 $\\
			&$16647-16i $& $ 88\% $&$ 12\% $&$ 0.56 $&$ 0.64 $&$ 0.34 $&$ 0.50 $&$ 0.67 $&$ 0.56 $\\
			\hline\hline
		\end{tabular*}
	\end{table*}
	
	The complex eigenenergies of the $ cc\bar c\bar b $ and $ bb\bar b\bar c $ systems are shown in Fig.~\ref{fig:cccb} and Fig.~\ref{fig:bbbc}, respectively. We obtain a series of resonant states in these systems, whose complex energies, proportions of different color configurations and rms radii are summarized in Table~\ref{tab:cccb_structure} and Table~\ref{tab:bbbc_structure}. The $ cc\bar c\bar b $ resonant states lie within the mass region $ (10.0,10.3) $ GeV, while the $ bb\bar b\bar c $ resonant states lie within the mass region $ (16.5,16.7) $ GeV.   The different rms radii of these states are of the same order, falling between the sizes of the corresponding 1S and 2S mesons. This indicates that all of these resonant states have compact tetraquark configuration. 
	
	We observe a great resemblance between the $ cc\bar c\bar b $ and $ bb\bar b\bar c $ systems. There are four resonant states with $ J^P=0^+ $, six resonant states with $ J^P=1^+ $ and three resonant states with $ J^P=2^+ $ in both systems. In each system, the masses of the resonant states with the same quantum number are very close to each other. The emergence of a large number of resonant states (especially in the $ 1^+ $ systems) may result from the coupling between numerous near-degenerate dimeson thresholds.
	
	In Ref.~\cite{Hu:2022zdh}, the authors adopted a different quark potential model to investigate the $ cc\bar c\bar b $ and $ bb\bar b\bar c $ systems and applied the real scaling method (RSM) to identify genuine resonant states. In each system, they found three resonant states with different quantum numbers. It is worth mentioning that the CSM and the RSM share similar physical principles, as the energy eigenvalues of bound and resonant states do not change under complex or real scaling transformations. Additionally, even with different quark potential models, the masses and widths of the lowest resonant states obtained in our calculations are in fair agreement with the results from Ref.~\cite{Hu:2022zdh}, as shown in Table~\ref{tab:cccb&bbbc}. The CSM makes it easier to identify resonant states and calculate their widths. Our calculations have identified more resonant states and obtained more stable width values compared with Ref.~\cite{Hu:2022zdh}. 

	\begin{table}
		\centering
		\caption{The masses $ M $ and widths $ \Gamma $ (in MeV) of the lowest $ cc\bar c\bar b $ and $ bb\bar b\bar c $ resonant states in our calculations and in Ref.~\cite{Hu:2022zdh}.}
		\label{tab:cccb&bbbc}
		\setlength{\tabcolsep}{2.5mm}
		\label{tab:cccb&bbbc}
		\begin{tabular}{cc|cc|cc}
			\hline\hline
			Systems & $J^{P}$ & \multicolumn{2}{c|}{This work} & \multicolumn{2}{c}{Ref.~\cite{Hu:2022zdh}} \\
			&  & $M$         & $\Gamma$         & $M$         & $\Gamma$        \\
			\hline
			$ cc\bar c\bar b $&$ 0^{+} $ &$ 10059 $&$ 4 $ &$ 10079 $&$ 6.7\sim8.4 $\\
			&$ 1^{+} $& $ 10068 $&$ 1 $ &$ 10081 $&$ 1.4\sim7.2 $\\
			&$ 2^{+} $& $ 10169 $&$ 20 $ &$ 10177 $&$ 9.1\sim11.1 $\\
			\hline 
			$ bb\bar b\bar c $&$ 0^{+} $ &$ 16511 $&$ 10 $ &$ 16474 $&$ 2.2\sim6.1 $\\
			&$ 1^{+} $&$ 16515 $&$ 2 $ &$ 16474 $&$ 2.2\sim 6.9 $\\
			&$ 2^{+} $&$ 16554 $&$ 14 $ &$ 16541 $&$ 5.3\sim 8.5 $\\
			\hline \hline 
		\end{tabular}
	\end{table}

	\section{Summary}\label{sec:summary}
	In summary, we calculate the mass specturm of the S-wave fully heavy tetraquark systems with different flavors $ (bc\bar b\bar c,bb\bar c\bar c, cc\bar c\bar b, bb\bar b\bar c) $ using the AP1 quark potential model, which was also adopted to study the fully charmed and fully bottomed tetraquark systems in our previous study~\cite{Wu:2024euj}. We apply the complex scaling method to  identify genuine resonant states from meson-meson scattering states, and the Gaussian expansion method to solve the four-body Schrödinger equation.
	
	We obtain a series of resonant states in all these systems with different quantum numbers. Specifically, the $ bc\bar b\bar c, bb\bar c\bar c, cc\bar c\bar b, bb\bar b\bar c $ states are predicted to lie within the mass regions of $ (13.2,13.5) $, $ (13.3,13.6) $, $ (10.0,10.3) $, $ (16.5,16.7) $ GeV, respectively. They all lie above the $M(1S)M'(2S)$ dimeson thresholds, with two-body strong decay widths ranging from less than $ 1 $ MeV to around $ 70 $ MeV. Among these states, the $ bc\bar b\bar c $ tetraquark states may be the most promising ones to be discovered experimentally in the near future. We recommend experimental exploration of the $ 1^{++} $ and $ 2^{++} $ $ bc\bar b\bar c $ states in the $ J/\psi\Upsilon $ channel, including the $ T_{bc\bar b\bar c,1^{++}}(13255)$, $T_{bc\bar b\bar c,1^{++}}(13276)$, $T_{bc\bar b\bar c,1^{++}}(13310)$, $T_{bc\bar b\bar c,1^{++}}(13318)$, $T_{bc\bar b\bar c,1^{++}}(13355)$ and $T_{bc\bar b\bar c,2^{++}}(13333) $.
	
	We calculate the root-mean-square (rms) radii to analyze the spatial structures of the tetraquark states. We find that all fully heavy tetraquark resonant states with different flavors obtained in our calculations are compact tetraquark states. Moreover, we improve the rms radii results of the fully charmed tetraquark states in our previous work~\cite{Wu:2024euj}. With the improved results, we reidentify three fully charmed tetraquark resonant states as compact tetraquark states. As a result, we find that all fully heavy tetraquark states in our calculations have compact tetraquark configuration.

	\section*{ACKNOWLEDGMENTS}
	
	We thank Zi-Yang Lin for the helpful discussions. This project was supported by the National
	Natural Science Foundation of China (11975033 and 12070131001). This
	project was also funded by the Deutsche Forschungsgemeinschaft (DFG,
	German Research Foundation, Project ID 196253076-TRR 110). The computational resources were supported by High-performance Computing Platform of Peking University.
	
	\newpage
	\bibliography{QQQQ'ref}
	
	\onecolumngrid
	\clearpage
	\twocolumngrid

\end{document}